\documentstyle[preprint,aps]{revtex}
\def\ie{ {i.e.,} }

\def\d{{\rm d}}

\def\dt{{\delta \theta}}

\def\dt2{(\delta \theta)^2}

\def\p{\sigma}
\def\al{\alpha}

\def\im{{\rm i}}

\def\lan{\left\langle}
\def\ran{\right\rangle}

\def\ekp{\epsilon_{kp}}
\def\delx{\partial_x}

\def\nonum{\nonumber}

\def\dl{{{{\rm d}} \over {{{\rm d}} \ell}}}
\def\dt{\frac{\partial}{\partial T}}

\def\e{{\rm e}}

\def\titheta{\tilde{\theta}}

\def\tiphi{\tilde{\phi}}

\def \virg{\;\;,}
\def \point{\;\,.}
\def \e{{\rm e }}
\def\jo #1#2#3#4{#1 {\bf #2}, #4  (#3)}  

\def\PRB{Phys.\ Rev.\ B}
\def\PRL{Phys.\ Rev.\ Lett.}

\def\JPC{J.\ Phys.\ C}

\def\JPSJ{J.\ Phys.\ Soc.\ Jpn.}

\def\ADV{Adv.\ Phys.}

\def\PHC{Physica C}

\def\MCLC{Mol.\ Cryst.\ Liq.\ Cryst.}
\begin{document}
\draft
\title
{
Spin-density wave versus superconducting fluctuations \\
for quasi-one-dimensional electrons \\ 
in two chains of Tomonaga-Luttinger liquids
}

\author
{
Hideo Yoshioka
and Yoshikazu Suzumura \\
}

\address
{
Department of Physics, Nagoya University, \\
Nagoya 464-01, Japan
} 

\date{Received 1 March 1996}

\maketitle

\begin{abstract}
 We study possible states at  low temperatures  
 by applying the  renormalization-group method   
 to two chains of Tomonaga-Luttinger liquids 
 with both  repulsive intrachain interactions 
 and  interchain hopping. 
   As the energy  decreases below the hopping energy, 
   three distinct  regions I, III,  and II appear  successively 
   depending on properties of fluctuations.  
 The crossover from the spin-density wave (SDW) state 
 to superconducting (SC) state takes place in region III 
  where   there are the excitation gaps 
 of  transverse charge and spin fluctuations. 
 The competition between SDW and SC states in region III  
   is crucial to  understanding 
 the phase  diagram  in the quasi-one-dimensional organic conductors.  
\end{abstract}


\narrowtext
 Organic conductors, $\rm (TMTSF)_2 X$ 
 are  typical quasi-one-dimensional (1D) electron systems 
 with the   ratio   given by  $t_a : t_b : t_c \simeq 10 : 1 : 0.1$  
 where  $t_a$, $t_b$ and $t_c$ are  the transfer energies 
 along the $a$-axis,   the $b$-axis  and the $c$-axis, respectively. 
  Such an anisotropy  gives  much varieties of  
 ordered states  characteristic in low dimension which have 
 been obtained  by varying temperature, pressure and magnetic field. 
 These conductors show  a common feature 
 of a phase diagram on the plain of pressure and 
 temperature\cite{Jerome-Schulz,Ishiguro-Yamaji}.  
  For example, ${\rm (TMTSF)_2 PF_6}$ undergoes a transition from 
 the metallic state to the insulating state of
  spin-density wave (SDW)   at  about 20 K in ambient pressure. 
  Under pressure,  the transition temperature of the SDW decreases 
  and a  superconducting (SC) phase appears 
 at a critical temperature of 0.9 K for 12 kbar\cite{Jerome}. 
 Such a competition between the SDW state and the SC state has been 
  ascribed to 
  the fact that the nesting condition for the Fermi surface 
 is essential for the formation of the SDW state  
 but not for the  SC state. 
 Actually the increase of pressure 
 deforms the 1D band  into the three-dimensional (3D)  one, 
 and thus leads to breaking  of the nesting condition.  
 There are many  theoretical work on the problem of the competition of 
 the SC state and the SDW state in the organic conductors, 
 e.g., 
 the phase diagram of the quasi-1D electron systems with the interchain hopping  
 has been  studied by use of the g-ology theory
 where  the renormalization of the  1D 
 fluctuation stops at the energy of the interchain hopping\cite{Psaltakis}.

  Two chains of Tomonaga-Luttinger liquids coupled by interchain 
 hopping is a basic model for understanding 
  the phase diagram  of the quasi-1D systems.  
  Extensive studies on the  two chain systems 
 have shown that  the ground state 
 in the case of the  weak repulsive intrachain interactions 
is given by the SC state 
 with interchain and in phase  pairing
\cite{Fabrizio,Yamaji-Shimoi,Nagaosa-Oshikawa,Schulz,Balents-Fisher}. 
 However, the  conclusion derived from these  two coupled chains is apparently 
 incompatible  with the phase diagram of 
 the above quasi-1D conductors due to the following fact.  
 The reason for the SDW state  in the phase diagram is considered as 
 the good nesting condition at low pressure, 
 while  the two  chains with the perfect nesting condition 
 shows the SC state instead of the SDW state.  
 In the present paper,  we examine the two chain systems 
 in terms of the renormalization-group method 
 to answer such a problem. 
 It is shown that the phase transition from the SDW to the SC state 
 under the pressure can be understood qualitatively in the two chain system.   

 We consider the Hamiltonian of the two chains  written as 
\begin{eqnarray}
{\cal H} &=& 
\sum_{k,p,\p,i} \ekp a^{\dagger}_{k,p,\p,i} a_{k,p,\p,i}  
- t \sum_{k,p,\p} \left\{ a^{\dagger}_{k,p,\p,1} a_{k,p,\p,2} + ( 1 
\leftrightarrow 2 ) \right\} 
\nonum \\
&+&
{ {\pi v_F } \over {L} } g_1 \sum_{p,\p,\p',i} \sum_{k_1, k_2, q}
a^{\dagger}_{k_1,p,\p,i} a^{\dagger}_{k_2,-p,\p',i} a_{k_2+q,p,\p',i} 
a_{k_1-q,-p,\p,i} \nonum \\
&+& 
{ {\pi v_F } \over {L} } g_2 \sum_{p,\p,\p',i} \sum_{k_1, k_2, q}
a^{\dagger}_{k_1,p,\p,i} a^{\dagger}_{k_2,-p,\p',i} a_{k_2+q,-p,\p',i} 
a_{k_1-q,p,\p,i} \point  
\label{eqn:hf} 
\end{eqnarray}
 The quantity  $k$ and $\ekp = v_F(pk-k_F)$  are 
 the momentum and the kinetic energy of a Fermion   
 where  $v_F$, $p = +(-)$ and $k_F$ denote  the Fermi velocity, 
 the right-going (left-going) state of an electron and 
 the Fermi momentum, respectively. 
 A creation operator of the Fermion with indices, 
$k,p,\p,i$ 
is expressed by $a^{\dagger}_{k,p,\p,i}$ where   
$\p = +(-)$ and  $i(=1,2) $ correspond to  
the spin $\uparrow(\downarrow)$ state and the index of the chains.  
 The interchain hopping  and the length of the chain are 
 defined  by $t$ and $L$. 
 The normalized quantities  
 $g_{1}$ and $g_{2}$,    
 respectively, express the matrix elements of the intrachain interaction for 
the backward scattering  and the forward scattering, 
  where the conventional definition of the elements are 
 given by $g \to g /(2 \pi v_F)$\cite{Solyom}. 
 We examine the case that the interactions are 
 repulsive, \ie $g_1$ and $g_2$ are positive values. 

  In Eq.(\ref{eqn:hf}), 
  we  take account of the splitting of the Fermi surface  
 due to the interchain hopping. 
  Then, by utilizing the method of Abelian bosonization to the new Fermi  
surface,  
  we express the  Hamiltonian  in terms of  the phase variables 
  as   
\begin{eqnarray}
	{\cal H} & = & \frac{v_F}{4\pi} \int \d x 
	\Big\{ (1 + g_\theta) (\delx \theta_+)^2 + (1 - g_\theta) (\delx 
\theta_-)^2 \Big\} \nonum \\
	& + & \frac{v_F}{4\pi} \int \d x 
	\left\{ (1 -g_1 ) (\delx \phi_+)^2 + (1 + g_1 ) (\delx \phi_-)^2 
\right\} \nonum \\
	& + & \frac{v_F}{4\pi} \int \d x 
	\left\{ (\delx \titheta_+)^2 + (\delx \titheta_-)^2 
          + (\delx \tiphi_+)^2 + (\delx \tiphi_-)^2 \right\} \nonum \\
	& + & \frac{v_F}{2 \pi \al^2} g_\theta
	\int \d x \left\{ 
	\cos \sqrt{2} \titheta_- - \cos (2 q_0 x - \sqrt{2} \titheta_+) \right\} 
    \left\{ \cos \sqrt{2} \tiphi_- + \cos \sqrt{2} \tiphi_+ \right\} 
	\nonum \\
	& - & \frac{v_F}{2 \pi \al^2} g_1
	\int \d x \left\{ 
	\cos \sqrt{2} \titheta_- + \cos (2 q_0 x - \sqrt{2} \titheta_+) \right\}
	\left\{ \cos \sqrt{2} \tiphi_- - \cos \sqrt{2} \tiphi_+ \right\} 
	\nonum \\
	& + & \frac{v_F}{\pi \al^2} g_{1}
	\int \d x \cos \sqrt{2} \phi_+ 
	\Big\{ \cos \sqrt{2} \tiphi_+ + \cos \sqrt{2} \tiphi_- 
	 - \cos \sqrt{2} \titheta_- +  \cos (2 q_0 x - \sqrt{2} \titheta_-) 
	\Big\} \virg  
	\label{eq:ht}  
\end{eqnarray}
where 
 $g_\theta \equiv 2 g_{2} - g_{1}$ 
 and $q_0 = 2 t /v_F$.   
  The quantity  $\al^{-1}$ is the upper cutoff  of 
 the wave number. 
  In Eq.(\ref{eq:ht}), the phase fields   are defined by
\begin{eqnarray}
& & \sum_{q \neq 0} \frac{2 \pi \im}{q L}  e^{- \al |q|/2 - \im q x}
a^\dagger_{k+q,p,\p,\mu} a_{k,p,\p,\mu} \nonum \\
&=& \frac{1}{2\sqrt{2}}
\left\{ 
\theta_+ + p \theta_-
+ \mu (\titheta_+ + p \titheta_-)
+ \p (\phi_+ + p \phi_-)
+ \p \mu  (\tiphi_+ + p \tiphi_-)
\right\} \virg 
\label{eq:phase}
\end{eqnarray}
where $a_{k,p,\p,\mu} = ( -\mu a_{k,p,\p,1} + a_{k,p,\p,2} ) / \sqrt{2}$ 
($\mu = \pm$). 
 The phase variables, $\theta_\pm$ and $\phi_\pm$, 
  describe the fluctuations of the total charge and 
 the total spin, respectively, while 
 $\titheta_\pm$ and $\tiphi_\pm$ 
 express the transverse fluctuation of the charge and 
 the spin degrees of freedom. 
 In terms of Eq.(\ref{eq:phase}), the field operators of Fermions, 
 $\psi_{p,\p,\mu}$ ($= 1/\sqrt{L} \sum_k \e^{\im k x} a_{k,p,\p,\mu}$),  
 are expressed as \cite{Luther-Peschel,Luther-Emery}   
\begin{eqnarray}
 \psi_{p,\p,\mu} ( x ) &=& { 1 \over \sqrt{2 \pi \al} }
 \exp \left[ 
{\rm i} p k_{F\mu} x + \im \Theta_{p,\p,\mu} + \im \pi \Xi_{p, \p, \mu} 
         \right] \virg 
\label{eq:field-1} 
\end{eqnarray}
with 
\begin{equation}
\Theta_{p,\p,\mu} = \frac{1}{2\sqrt{2}}
\Big\{ 
  p \theta_+ + \theta_- + \mu (  p \titheta_+ + \titheta_- )
+ \p ( p \phi_+ + \phi_- ) + \p \mu (  p \tiphi_+ + \tiphi_- )
\Big\} \point 
\label{eq:field-2}
\end{equation}
 In Eq.(\ref{eq:field-1}), the phase factor, $\pi \Xi_{p, \p, \mu}$, 
 is added so that the Fermion operators with different indices 
satisfy the anticommutation relation\cite{Solyom}.   
 The factor $ \Xi_{p, \p, \mu}$ is given by 
 $\Xi_1 =0$ and $\Xi_i = \sum_{j=1}^{i-1} \hat N_j$, $(i = 2 \sim 8)$ 
 with ${\hat N}_i$ being the number operator of the Fermions with indices $i$  
 where  the index $(p,\p,\mu)$ corresponds to  
$(+,+,+) = 1$, $(+,-,+) = 2$, $(+,+,-) = 3$, $(+,-,-) = 4$, 
$(-,+,+) = 5$, $(-,-,+) = 6$, $(-,+,-) = 7$ and $(-,-,-) = 8$,   
respectively. 
 
 In the case of $g_{2} \neq 0$ and $g_{1} = 0$, 
 it has been shown that 
 the  fourth term in Eq.(\ref{eq:ht}) without the term including the 
 misfit parameter 
 tends to the strong coupling in the low-energy 
limit\cite{Finkelstein-Larkin}. 
 In this case, since  the transverse fluctuation of the charge becomes 
 completely  gapful and that of the  spin shows two kinds of 
 excitations being gapless and gapful, 
  the possible states are given by 
 the density wave with  both intrachain  and out of phase ordering and 
 the SC states with both interchain  and in phase pairing
\cite{Schulz}.  
 By use of Eqs.(\ref{eq:field-1}) and (\ref{eq:field-2}), 
 those states, $DW_+^{\p,\p'}$ for the density wave and 
 $S_-^{\p,\p'}$ for the SC states, are expressed explicitly 
 in terms of the phase variables as follows\cite{Yoshioka-Suzumura-4},        
\begin{eqnarray}
DW_{+}^{\p,\p'} &\equiv& 
\sum_{\mu} \psi^\dagger_{p,\p,\mu} \psi_{-p,\p',-\mu} 
 =  - \left\{ \psi^\dagger_{p,\p,1} \psi_{-p,\p',1} - (1 \to 2)  \right\}
\nonum \\
& \sim & 
\frac{-\im}{\pi \al}
\e^{-\im 2 p k_F x}  
\e^{\frac{-\im p}{\sqrt{2}}\theta_+}
\e^{\frac{-\im p}{2\sqrt{2}}(\p + \p')\phi_+} 
\e^{\frac{-\im}{2\sqrt{2}}(\p - \p')\phi_-} \nonum \\ 
 & \times &
 \sin \left\{ \frac{\titheta_-}{\sqrt{2}} + p 
\frac{\p-\p'}{2\sqrt{2}}\tiphi_+ 
+ \frac{\p +\p'}{2\sqrt{2}}\tiphi_- \right\}, 
\label{eq:DW+} \\
S_{-}^{\p,\p'} &\equiv& 
\sum_{\mu} \mu \psi_{p,\p,\mu} \psi_{-p,\p',\mu}  
 =  - \left\{ \psi_{p,\p,1} \psi_{-p,\p',2} + (1 \leftrightarrow 2) 
\right\}
\nonum \\
 & \sim & 
\frac{\im}{\pi \al} 
\e^{\frac{\im}{\sqrt{2}}\theta_-}
\e^{\frac{\im p}{2\sqrt{2}}(\p- \p')\phi_+} 
\e^{\frac{\im}{2\sqrt{2}}(\p + \p')\phi_-} \nonumber \\ 
 & \times & 
 \sin \left\{ \frac{\titheta_-}{\sqrt{2}} + p 
\frac{\p-\p'}{2\sqrt{2}}\tiphi_+ 
+ \frac{\p +\p'}{2\sqrt{2}}\tiphi_- \right\},  
\label{eq:S-} 
\end{eqnarray}
where $\psi_{p,\p,\mu} = 1 / \sqrt{L} \sum_k a_{k,p,\p,\mu} \e^{\im k 
x}$.  
  Among two kinds of states,  
  $DW_{+}^{\p,\p'}$ becomes  dominant state 
 and $S_{-}^{\p,\p'}$ is subdominant\cite{Schulz}. 
 In the following, we discuss the effects of the backward scattering 
 on the the competition between these state and show 
 that the crossover from the SDW state to the SC state are seen when the 
 scaling energy are decreased.       

 In Eq.(\ref{eq:ht}),  
   the interchain hopping leads to two kinds of energy regions, \ie  
 the high-energy region  
 where the excitations are  essentially the 
 same as those in the absence of the hopping 
 and the low-energy region 
 where the hopping is relevant. 
  By applying  the renormalization procedure of the  1D system 
  to the higher-energy region($\omega > t$)\cite{Schulz}, 
 the  effective Hamiltonian in the case of  $\omega < t$ 
 is given by      
\begin{eqnarray}
	{\cal H} & = & \frac{v_\theta}{4\pi} \int \d x 
	\left\{ \frac{1}{\eta_\theta} (\delx \theta_+)^2 
	      + \eta_\theta (\delx \theta_-)^2 \right\} 
	 +  \frac{v_\phi}{4\pi} \int \d x 
	\left\{ \frac{1}{\eta_\phi} (\delx \phi_+)^2 
	      + \eta_\phi (\delx \phi_-)^2 \right\} \nonum \\
	& + & \frac{v_F}{4\pi} \int \d x 
	\left\{ \frac{1}{\eta_{\titheta}} (\delx \titheta_+)^2 
	      + {\eta_{\titheta}} (\delx \titheta_-)^2 \right\} 
	 +  \frac{v_F}{4\pi} \int \d x	       
    \left\{ \frac{1}{\eta_{\tiphi}} (\delx \tiphi_+)^2 
          + {\eta_{\tiphi}} (\delx \tiphi_-)^2 \right\} \nonum \\ 
	& + & \frac{v_F}{\pi \al'^2} g_-
	      \int \d x \cos \sqrt{2} \titheta_-  \cos \sqrt{2} \tiphi_- 
     +  \frac{v_F}{\pi \al'^2} g_+
	      \int \d x \cos \sqrt{2} \titheta_-  \cos \sqrt{2} \tiphi_+ \nonum 
\\ 	
	& + & \frac{v_F}{\pi \al'^2} g^*_a
	      \int \d x \cos \sqrt{2} \phi_+  \cos \sqrt{2} \tiphi_+ 
     +  \frac{v_F}{\pi \al'^2} g^*_b
	      \int \d x \cos \sqrt{2} \phi_+  \cos \sqrt{2} \tiphi_- \nonum \\ 
	& - & \frac{v_F}{\pi \al'^2} g^*_c
	      \int \d x \cos \sqrt{2} \phi_+  \cos \sqrt{2} \titheta_-                             \virg   
	\label{eq:hl}  
\end{eqnarray}  
 where $\al' \sim v_F / t >  \al$, 
 $v_\theta = v_F \sqrt{( 1 + g_\theta )( 1 - g_\theta )}$,  
 $v_\phi = v_F \sqrt{( 1 - g^*_1 )( 1 + g^*_1 )}$, 
 $\eta_\theta = \sqrt{( 1 - g_\theta )/( 1 + g_\theta )}$, 
 $\eta_\phi = \sqrt{( 1 + g^*_1 )/( 1 - g^*_1 )}$, 
 $\eta_{\titheta} = \eta_{\tiphi} = 1$,  
 $g_\pm = g_2 - g_1/2 \pm g^*_{1}/2$, and 
 $g^*_a = g^*_b = g^*_c = g^*_{1}$.  
 The quantity  
 $g^*_{1}$
 is the  renormalized 
 matrix element of the backward scattering and given by 
 $ g^*_1  \equiv g_1/\big\{ 1 + 2 g_1 \ln (v_F/t\al) 
 \big\} $\cite{Solyom}.   
  In this case, the symmetry of spin degree of freedom between  
 $\phi_+$ and $\phi_-$, and that between $\tiphi_+$ and  $\tiphi_-$ 
 are broken due to the finite magnitude of 
 the renormalized  backward scattering\cite{Schulz} 
 which reduces to zero in the limit of $t \to 0$. 
 Hereafter we assume $g_2 > g_1/2$ and then $\eta_\theta < 1$.  
  We note  that the present $\eta_\theta$  corresponds to $K$ 
 in Ref. 8 and $1/K_{+\rho}$ in Ref. 9.   

 Properties at temperatures lower than the hopping energy 
 are examined by applying  the  renormalization-group method 
 to Eq.(\ref{eq:hl}). 
 By making use of the scaling relation for 
 correlation functions  
$\lan \exp \big[ - \im \titheta_+(x)/ \sqrt{2} \big] 
\exp \big[ \im \titheta_+(0)/ \sqrt{2} \big] \ran$, 
$\lan \exp \big[ - \im \tiphi_+(x)/ \sqrt{2} \big] 
\exp \big[ \im \tiphi_+(0)/ \sqrt{2} \big] \ran $, and  
$\lan \exp \big[ - \im \phi_+(x)/ \sqrt{2} \big] 
\exp \big[ \im \phi_+(0)/ \sqrt{2} \big] \ran$\cite{Giamarchi-Schulz},  
 the equations of the renormalization-group are obtained  as   
\begin{eqnarray}
	\dl \eta_{\titheta} & = & \frac{1}{4} ( g_-^2 + g_+^2 + g^{*2}_c), 
	\label{eq:rg1} \\
	\dl \eta_{\tiphi} & = & \frac{1}{4} ( g_-^2 - g_+^2 \eta_{\tiphi}^2 
	                                    - g^{*2}_a \eta_{\tiphi}^2 + 
g^{*2}_b), 
	\label{eq:rg2} \\
	\dl \eta_{\phi} & = & - \frac{\eta_\phi^2}{4} ( g^{*2}_a + g^{*2}_b + 
g^{*2}_c ), 
	\label{eq:rg3} \\
	\dl g_- & = & ( 2 - \frac{1}{\eta_{\titheta}} - \frac{1}{\eta_{\tiphi}} ) 
g_-,
	\label{eq:rg4} \\
	\dl g_+ & = & ( 2 - \frac{1}{\eta_{\titheta}} - \eta_{\tiphi} ) g_+,
	\label{eq:rg5} \\
	\dl g^*_a & = & ( 2 -\eta_{\phi} - \eta_{\tiphi} ) g^*_a,
	\label{eq:rg6} \\
	\dl g^*_b & = & ( 2 -\eta_{\phi} - \frac{1}{\eta_{\tiphi}} ) g^*_b,
	\label{eq:rg7} \\
	\dl g^*_c & = & ( 2 -\eta_{\phi} - \frac{1}{\eta_{\titheta}} ) g^*_c ,
	\label{eq:rg8}
\end{eqnarray}
  where  $\ell \equiv {\rm ln} (t/\omega)$.  
The initial conditions are given by 
 those at $\ell = 0$,  i.e., 
 $\eta_\phi(0) = \sqrt{( 1 + g^*_1 )/( 1 - g^*_1 )}$ $(>1)$, 
 $\eta_{\titheta}(0) = \eta_{\tiphi}(0) = 1$,  
 $g_\pm(0) = g_2-g_1/2 \pm g^*_1/2$,  
 and  $g^*_a(0) = g^*_b(0) = g^*_c(0) = g^*_1$, respectively. 
 
  We numerically calculate  the differential equations 
  (\ref{eq:rg1})$-$(\ref{eq:rg8}). 
 The result in the  case of $g_1 = 0.45$, $g_2 = 0.5$  and $t \al / v_F = 
0.1$  
 is  shown in Fig.\ref{fig:rg}. 
 There are three kinds of regions,
$t > \omega >\omega_1$ (I), 
$\omega_1 > \omega >\omega_2$ (III) and 
$\omega_2 > \omega$ (II) 
 where $\ln (t/\omega_1) \simeq 8$ and  $\ln (t/\omega_2) \simeq 10$. 
 The corresponding regions are marked along the horizontal axis. 

 In region I, 
 the monotonical increase of $\eta_{\titheta}$ 
   and decrease of $\eta_{\tiphi}$ are obtained  
  since $g_+$ increases  toward the  strong coupling and   
  the quantity $g_-$ with a slight enhancement decreases  to zero.  
  The variations of 
 the nonlinear terms including  $\phi_+$, and $\eta_\phi$ are negligible.  

  In region III,   the relevant $g_+$ results in  
  the gaps for the transverse fluctuations of both 
 charge and spin degrees of freedom. 
 Note that, in the present case, there is no gapless mode  of the 
 transverse spin fluctuation which exists in the case of only the forward 
 scattering.   
 In this region, 
 the variables $\titheta_-$ and $\tiphi_+$ show the relevance 
 and then the pairing states with antiparallel spin 
 in Eqs.(\ref{eq:DW+}) and (\ref{eq:S-}),  \ie 
 the transverse spin-density wave (TSDW) state  
 and the singlet superconducting (SS) state are selected. 
 Since the exponent  of the correlation function  
 for the TSDW (SS) state is given by 
 $(\eta_\theta + \eta_{\phi}^{-1})/2$ 
  ( $(\eta_\theta^{-1} + \eta_{\phi})/2$ ) and $\eta_\phi$ decreases to 
  zero, 
 the TSDW (SS) state  becomes  the most dominant  
 and the SS (TSDW) state  remains  subdominant 
 for $\omega$ near $\omega_1$ ($\omega_2$). 
 Therefore there is a crossover from the TSDW state to the SS state 
 in region III.   
 
  In region II, 
   the gap for the fluctuation  of the total spin  also appears 
  in addition to  the above transverse degrees of freedom. 
 In this region,  the nonmagnetic state is selected 
  because the gap of the total spin denotes  the finite amount 
 of energy for creating  the magnetization along the quantized axis.
 Actually the SS state  
 becomes 
dominant\cite{Fabrizio,Yamaji-Shimoi,Nagaosa-Oshikawa,Schulz,Balents-Fisher}, 
 while the correlation function of  TSDW  decays exponentially.  

 The existence of region III 
  is crucial  to relating the result of the two chains with 
 the phase diagram of quasi-1D conductors 
  as a function of temperature and pressure. 
  When the 3D coupling is taken into account,   
  the crossover from the TSDW state to the SS state 
 in region III of Fig.\ref{fig:rg}  
 gives rise to the actual phase  transition
\cite{Psaltakis}.  
   We assume  that the  pressure increases $t$  
 and then   the ratio $t/\omega$   
  increases with the fixed $\omega$.   
  Then  the crossover from  the fluctuation of TSDW to  that of SS   
  with the fixed $\omega$ 
 in Fig.1  indicates the transition from the SDW state to the SS state 
 in the phase diagram of organic conductors with the fixed $T$ 
\cite{Jerome-Schulz,Ishiguro-Yamaji}. 
 The SDW  in the two chains have nesting vector 
 $(2k_F, \pi)$, which satisfies the condition of the perfect nesting. 
 This result  is consistent with  
 the nesting vector obtained by NMR measurements
\cite{Delrieu,Takahashi}. 
 The  SS state  with  the interchain pairing suggests 
 the anisotropic gap which was asserted by the theoretical analysis
\cite{Hasegawa-Fukuyama} of the NMR relaxation rate of ${\rm (TMTSF)_2 
ClO_4}$  
\cite{Takigawa-Yasuoka-Saito}.   
Thus we obtained the correspondence between the states in region III 
and those in quasi-1D conductors. 

The present crossover from the SDW state to the SC state 
is due to the increase of $t$, \ie the dimensionality. 
The pressure also leads to the breaking of the nesting condition. 
This may cause an additional effect of the increase of the crossover energy.  

In conclusion, 
 we investigated electronic states  of 
  two chains of Tomonaga-Luttinger liquids  at  low temperatures 
 by use of the renormalization-group method. 
 There exist three kinds  of energy regions 
 where the crossover from the TSDW state  
 to the SS state occurs in region III. 
  The existence of such a crossover  is essential 
 to explaining the phase diagram of SDW vs SS states in  
 the quasi-1D conductors 
 such as ${\rm (TMTSF)_2 X}$. 

\vspace{1.0cm}

 One of the authors (H.Y) thanks K. Sano and N. Tanemura 
 for useful discussion. 
This work was financially supported  by a Grant-in-Aid for 
 Scientific Research from the Ministry of Education, 
 Science and Culture (No.05640410), 
 and  a Grant-in-Aid for Scientific  Research on the priority area,
 Novel Electronic States in Molecular Conductors from the 
 Ministry of Education,   Science and Culture, Japan.


\begin{figure}
\caption{
 Solutions of the renormalization-group equations of 
 Eqs.(9)$-$(16),   in the case of $g_1 = 0.45$, $g_2 = 0.5$,  
 and $t \alpha / v_F = 0.1$ 
 where  the boundary   between  regions I and  III (III and II)
 is given by 
 $\omega = \omega_1 \sim t {\rm e}^{-8}$ 
 ($\omega = \omega_2 \sim t {\rm e}^{-10}$).} 
\label{fig:rg}
\end{figure}

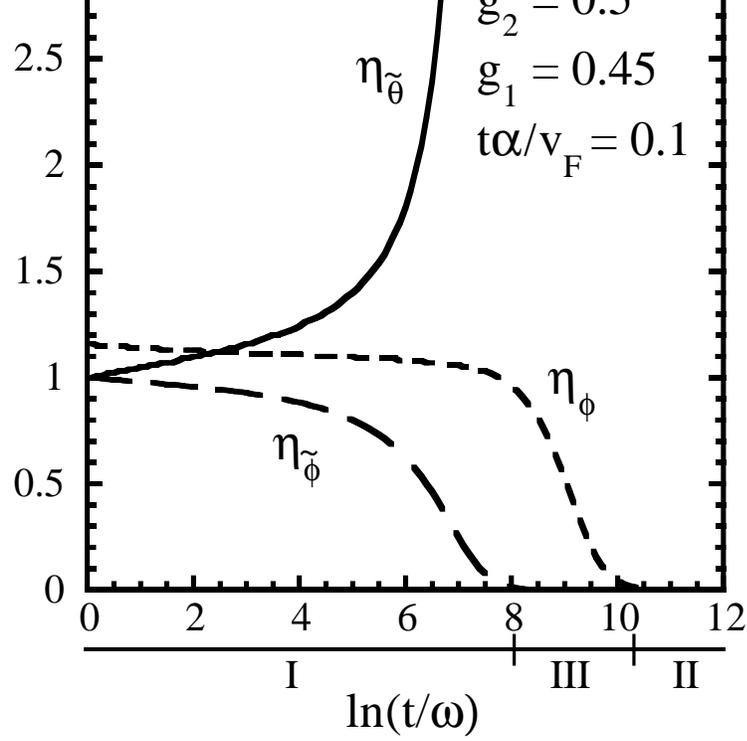
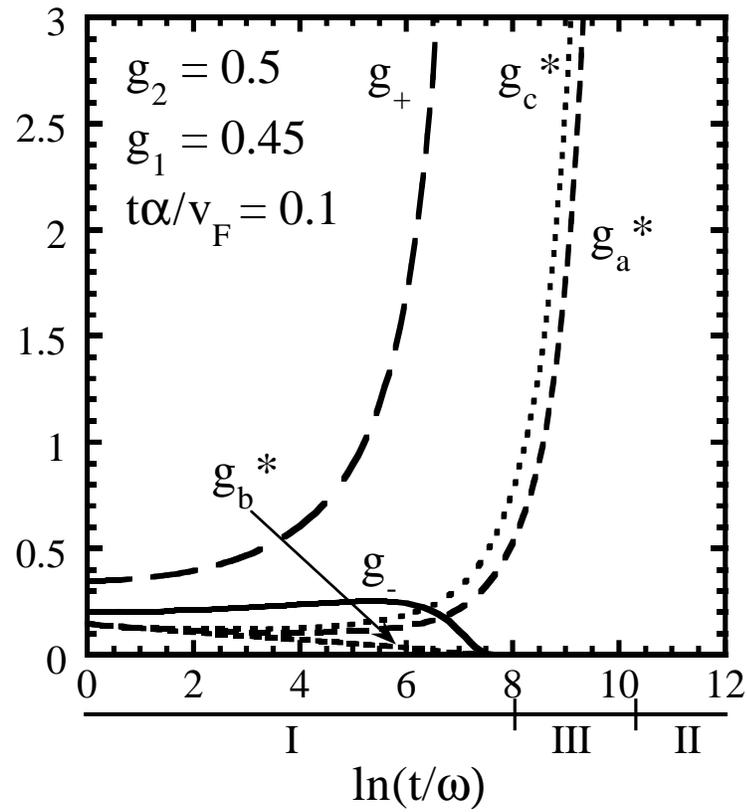
\end{document}